\documentclass[a4paper,11pt]{article}
\usepackage{epsfig}
\usepackage{amsmath}
\usepackage{fancyvrb}
\begin{document}
\large
\parskip = \baselineskip

  \newcounter{abc}
\renewcommand{\thefigure}{\arabic{figure}\alph{abc}}

\begin{center}
{\bf Rapid Equilibration by algorithmic quenching the ringing mode in
  molecular dynamics }

{G.J.Ackland}

{\normalsize
  School of Physics and Astronomy, The University of Edinburgh, James
  Clerk Maxwell Building, Kings Buildings, Mayfield Road,
  Edinburgh EH9 3JZ}
\end{center}

\subsection*{ABSTRACT}

Long wavelength acoustic phonons are normally weakly coupled to other
vibrational modes in a crystalline system.  This is particularly
problematic in molecular dynamics calculations where vibrations at the
system size scale are typically excited at initiation.  The
equilibration time for these vibrations depends on the strength of
coupling to other modes, so is typically very long.  A very simple
deterministic method which avoids the use of is presented which
removes this problem.

\subsection*{INTRODUCTION}

Molecular dynamics is a very powerful method for simulating the
behaviour of a many-atom system.  It is particularly useful and
flexible when used for studying phase transitions, since it makes no
assumption about the preferred crystal structure.  However if the
system is not initiated in an equilibrium structure, it may take some
time to equilibrate.  There is no good theory which enables one to say
how long the equilibration will take.  

Indeed, within statistical mechanics, it is difficult to define what a
non-equilibrium state even means. Provided the molecular dynamics is
based on some underlying Lagrangian, any simulation will be initiated
in a valid microstate of the appropriate ensemble.  The reason why
this microstate is ``non-equilibrium'' lies outside the realm of
statistical mechanics.

Probably the only sensible way to address this is via the
equipartition theorum, which holds that each degree of freedom has, on
average, an equal amount of energy.  One might then argue that if a
single degree of freedom in a microstate has a {\it macroscopic}
amount of energy - i.e. a finite fraction of the total energy - then
the microstate is atypical and ``non-equilibrium''.  One can think
here of non-equilibrium in the cartesian coordinate system ${r_i}$,
e.g.  a radiation damage simulation where the primary knock-on atom
has keV of energy, or non-equilibrium in the normal modes ${u_i}$ a
simulation in which one phonon mode is massively excited.

Unfortunately, even this definition fails in general because of the
nonunique definition of degrees of freedom.  One can always recast the
degrees of freedom in any linear combination - including a linear
combination  ${\mathcal Q}_i$ chosen such that all the kinetic energy
is in one degree of freedom ${\mathcal q}_1$.  We can observe that
such a perverse choice would have entangled degrees of freedom very
far from the normal modes which describe a typical near-harmonic
crystal system.  As such, the energy in ${\mathcal q}_1$ would
disperse quickly into the other ${\mathcal q}_j$ modes.

In this paper, we adopt the following definition of a
``non-equilibrium microstate'': {\it a microstate in which one degree
  of freedom has a macroscopic energy, and retains that macroscopic
  energy for many iterations of the system dynamics}.

In this definition, the degree of freedom with macroscopic energy can
be regarded as a thermodynamic variable.  The specific degree of
freedom considered here will be the size of the MD supercell in
constant pressure molecular dynamics.

\subsection*{THEORY}

The most common method for implementing constant pressure molecular
dynamics is the Parrinello Rahman barostat

The Parrinello-Rahman method can be used in molecular dynamics to
describe the NPE or NPT ensemble.  It has a number of well-known
anomalies
including lack of
rotational invariance and missing cross terms in the derivatives,
however in the current work we consider the original method, which is
widely used in molecular dynamics packages.
 In this, nine fictitious
degrees of freedom are introduced corresponding to the components of
the vectors defining the supercell  ${\bf a,b,c}$.  Each degree of freedom then has
its own equation of motion, derived from the  Parrinello-Rahman Lagrangian

\[ L = \frac{1}{2}\sum_i m_i\dot{\bf x'}_ih'h\dot{\bf x}_i
- U_{coh}- P(\Omega-\Omega_0)
+ \frac{1}{2}W {\rm Tr}\dot{h}'\dot{h}
-   \frac{1}{2} {\rm Tr}[h_0^{-1}({\bf S}-P){h'}_0\Omega_0h'h.  \]

in which {\bf x} are the fractional coordinates within the supercell
and the 3$\times$3 ``boxmatrix'' $h = [{\bf a,b,c}]$ define the
simulation volume. The scalar $W$ is the equivalent of a mass
associated with the box degrees of freedom.  P is the external
hydrostatic pressure, {\bf S} is the external stress (assumed hydrostatic in teh original work).  A constant term
$\Omega$Tr$({\bf S}-P)$ is ignored.
 
The atomic positions {\bf $r_j$} are written as a product of
fractional coordinates 

\[  {\bf r_j} =  h{\bf x_i} \]

from which a strain matrix with respect to a reference structure $h_0$
is defined by

\[ \epsilon = \frac{1}{2}({h'_0}^{-1}h'hh_0^{-1}-1) \]

where prime denotes the transpose.

The boxmatrix introduces nine additional degrees of freedom, three
stretches, three shears and three rotations.  Equations of motion for
these degrees of freedom come from the stresses on the supercell and
an equivalent kinetic energy term from their time derivatives and the
fictitious mass.

The scalar $W$ is the equivalent of a mass associated with the box degrees of
freedom. It determines how rapidly the box changes shape in response
to stress and can be related loosely to elastic constants. Typically
it takes a value of similar order of magnitude to the sum of the
atomic masses. From all this analysis, the equation of motion for the
boxmatrix degrees of freedom is

\[ W\ddot{h} = ({\bf \pi}-P)\sigma -h[h_0^{-1}({\bf S}-P){h'}_0\Omega_0] \]

where ${\bf \pi}$ is the internal stress tensor from the kinetic energy
and virial and ${\sigma}$ is defined by:

\[ \sigma = \{ b \times c, c \times a, a \times c \}.\]

The practical difficulty with this, and other barostats, is that the
box degrees of freedom are typically coupled only weakly to the atomic
degrees of freedom.  Thus if equipartition of energy between box and
atoms is not established at the start of the simulation, then
equilibration times become very long. This is particularly annoying
because one is not generally interested in the box degrees of freedom.
They are simply used to enable the cell to adjust its shape, to ensure
that the simulation is properly hydrostatic, or to facilitate phase
transitions.  It is easy to start the box degrees of freedom with an
equipartitioned kinetic energy, however the potential energy stored in
the strain field is {\it macroscopic}, and not generally known - often
this is what one is trying to find.  Furthermore, the
Parrinello-Rahman method was designed to facilitate phase transitions,
but when a phase transition occurs in an MD simulation it is
associated with a macroscopic release of strain energy - all of which
goes into the boxmatrix degrees of freedom.

The excess energy in the box degrees of freedom manifests itself in
harmonic oscillations of the entire system: these are long wavelength
phonons.  Because they correspond to the acoustic degrees of freedom,
they are referred to as ringing modes.  The important insight here is
that NPT simulations are out of equilibrium in this well-defined way.

Now that the problem is well defined, we can seek a bespoke solution
to stop the ringing mode.  Critical damping would be ideal, but that
would require a priori knowledge of the ringing frequencies in order
to determine the damping coefficient.  An alternative is to use an
algorithmic rather than Lagrangian formulation.  We are trying to
remove both potential and kinetic energy from one mode. Removing
kinetic energy is easy, one simply sets $\dot{h_{ij}}=0$.  However, if
this is done on each step, the potential energy associated with the
ringing remains out of equilibrium.  The most efficient strategy is to
remove the kinetic energy at the point where the potential energy is
minimised.  This can be done for each component of $h$ independently
with a single IF statement:

\[ {\rm IF} (\dot{h_{ij}}.\ddot{h_{ij}} < 0 ) {\rm THEN} \dot{h_{ij}} = 0\]

In most practical cases, this is extremely effective, however there
are two problems.   

Applying the algorithm to a single off-diagonal component is
equivalent to applying a torque to the system which is undesireable.
This traces back to the fact that Parrinello-Rahman introduces nine
fictitious variables, but three of these are rotations which should
not affect the energy.  In most practical MD codes, net rotation of
the cell can occur due to rounding errors in the stress tensor, and
are eliminated automatically.  An alternative is to transform
$\dot{h_{ij}}$ and $\ddot{h_{ij}}$ into volume expansion, five shears
and three rotations, apply the algorithm only to the first six.

A second problem is that the algorithic removal of energy does not
correspond to any hamiltonian dynamics, and therefore does not obey
time-reversibility and conservation of the hamiltonian energy.  The
resolution to this depends on what the simulation is meant to be
doing.  Energy conservation can be restored by recaling the velocities
of all the individual atoms in the system to compensate for the lost
energy.  This is effectively introducing a strong, fictituous,
anisotropic coupling between box and atom degrees of freedom to
resolve the weak-coupling problem.  An alternative is to observe that
the ringing mode comes from assuming that the acoustic mode is
localised: as consequence of the periodic boundary condition
approximation.  If one relaxes the approximation and assumes that
sound wave can travel away from the explicitly-modelled region, then
the loss of energy from the boxmatrix modes makes perfect physical
sense.

\subsection*{RESULTS}

The algorithmic quench was implemented in NPTp ensemble calculations in
the open source molecular dynamics code MOLDY.

To demonstrate its efficacy two examples are considered.  In both
cases we use a many-body empirical potential, first in simple
equilibration of a large system of 16000 lithium atoms, and second for
a small system of 2000 sodium atoms undergoing a martensitic phase
transition.

Three quantities are monitored to detect three distinct forms of
equilibration:

1/ Equilibration with respect to external work, the system volume, 

2/ Equilibration with respect to the First Law,  energy.

3/ Equilibration with respect to the Second Law,  entropy.

The first two are normally used to determine equilibration, but
thermodynamic equilibrium is defined by point at which the system no
longer generates entropy.  The entropy is calculated by integrating
the Central equation of thermodynamics:

\[ S = \int \frac {dU}{T} + \int \frac{PdV}{T}  \]

where the integral is converted to a time integral using the chain rule:

\[ dU = \frac{dU}{dt}dt; \hspace{1in}  dV = \frac{dV}{dt}dt \]

with the changes in U,V calculated numerically over a 0.1ps window.

{\bf \underline{NPE Crystal Equilibration}}

We consider first the case of equilibration at 280K, where the bcc
structure is expected to be stable. The structure is set up with atoms
on the perfect lattice structure, with the lattice parameter
determined from static relaxation.  Temperature is initialized at
280K, and maintained through use of a Nose thermostat.  

Due to thermal expansion, the initial volume is too small, and Figure
1(a) shows that without quenching there is almost no sign of
attenuation of the ringing more.  By contrast, the algorithmic quench
kills the ringing at the equilibrium value, and subsequent
oscillations are too small to discern.

The energy of the non-fictitious modes fluctuates at double the
frequency of the volume, and the alternating peaks heights in the
unquenched case shows that the effect of ringing on the other modes is
anharmonic, despite which the removal of energy from the mode is very
slow.  Again, the algorithmic quench stabilizes the energy in the
system from its first application.  Although the quench continues to
remove energy from the fictitious modes of the system, such that the
Hamiltonian energy reduces, the effect on the atomic modes is
negligible and cannot even be discerned in the figure.

The entropy of the quenched system rises in the equilibration period
and is then stable.  In the unquenched case, we see the system entropy
increase in expansion phases, and 
reduces in contraction.  The overall
downward slope indicates that the ringing is creating entropy in the
surroundings, and again there is no sign of this abating.

{\bf \underline{Martensitic transition} }

In the second case, we consider phase transitions in sodium.  The
system is initiated in the ground state fcc structure at 10K.  It is
then heated by incrementing the thermostat target temperature at the
rate of 1.5ps/k.  The potential describes two transformations, fcc-bcc
and melting.  At this heating rate, typical of MD simulation, the
figure shows that with standard NPT ensemble the ringing mode is
barely suppressed and, moreover, at the phase transition it is
enhanced.  With the algorithmic quench, equilibration is reached quickly.

One expects some hysteresis in a martensitic transition, and it is
notable here that the ringing simulation transforms first on heating, at a lower
temperature.  The ringing presumably helps to overcome the kinetic
barrier to transformation.

\subsection*{CONCLUSIONS}
We have identified the ``ringing'' mode as the slowest-equilibrating
mode in NPE molecular dynamics calculations. the ringing mode is
excited whenever the initial configuration is not at equilibrium, and
again at any stage in the simulation where a phase transformation
occurs.  In practical applications, it is therefore unavoidable. Using
a simple algorithm to remove energy from the mode, we show that
equilibration can be achieved many orders of magnitude faster than
with standard Parrinello Rahman Lagrangian.  Although the algorithm
does not correspond to a Lagrangian, and lacks strict
time-reversibility, we show that the equilibrated systems have the
same energy and volume properties as the Parrinello-Rahman.

M.Parrinello and A.Rahman,  Phys.Rev.Lett. {\bf 45}, 1196 (1980).
\newline
Nose, S., 1984. J.Chem. Phys  {\bf 81} , 511 (1984).
Cleveland,C.L.,  J.Chem.Phys.{\bf 89 }4987 (1989)
\newline
Nose,S  and Klein,M Mol Phys{\bf 50}, 1055 (1983).
\newline
Wentzcovitch R.M. Phys.Rev.B {\bf 44}, 2358 (1991)
\newline
G.J.Ackland, K D'Mellow, S.L.Daraszewicz, D.J.Hepburn,
  M Uhrin and K.Stratford Comp.Phys.Comm., {\bf 182} 2587 (2011)
\newline
A    Nichol, G. J. Ackland, Phys.Rev.B,
  submitted http://arxiv.org/abs/1601.06701 (2016)
\newline
U Pinsook, G.J. Ackland
Phys Rev B  {\bf 59}, 13642 (1999)
\newline
S Han, LA Zepeda-Ruiz, GJ Ackland, R Car, DJ Srolovitz
J. Appl Phys.  {\bf  93 } 3328 (2003)
\newline

\begin{figure}[h]
 \centering
                \includegraphics[width=0.45\textwidth]{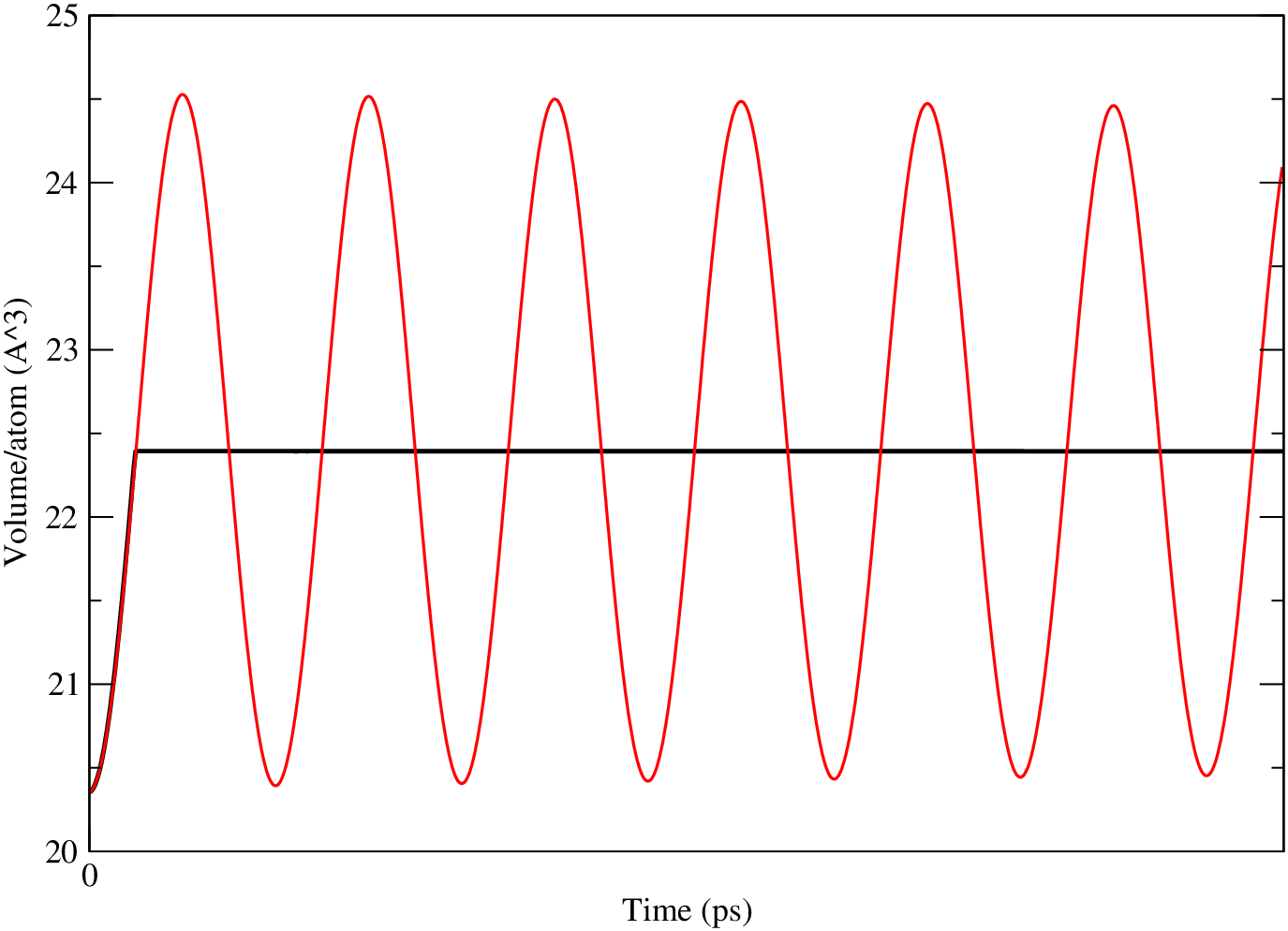}
                \includegraphics[width=0.45\textwidth]{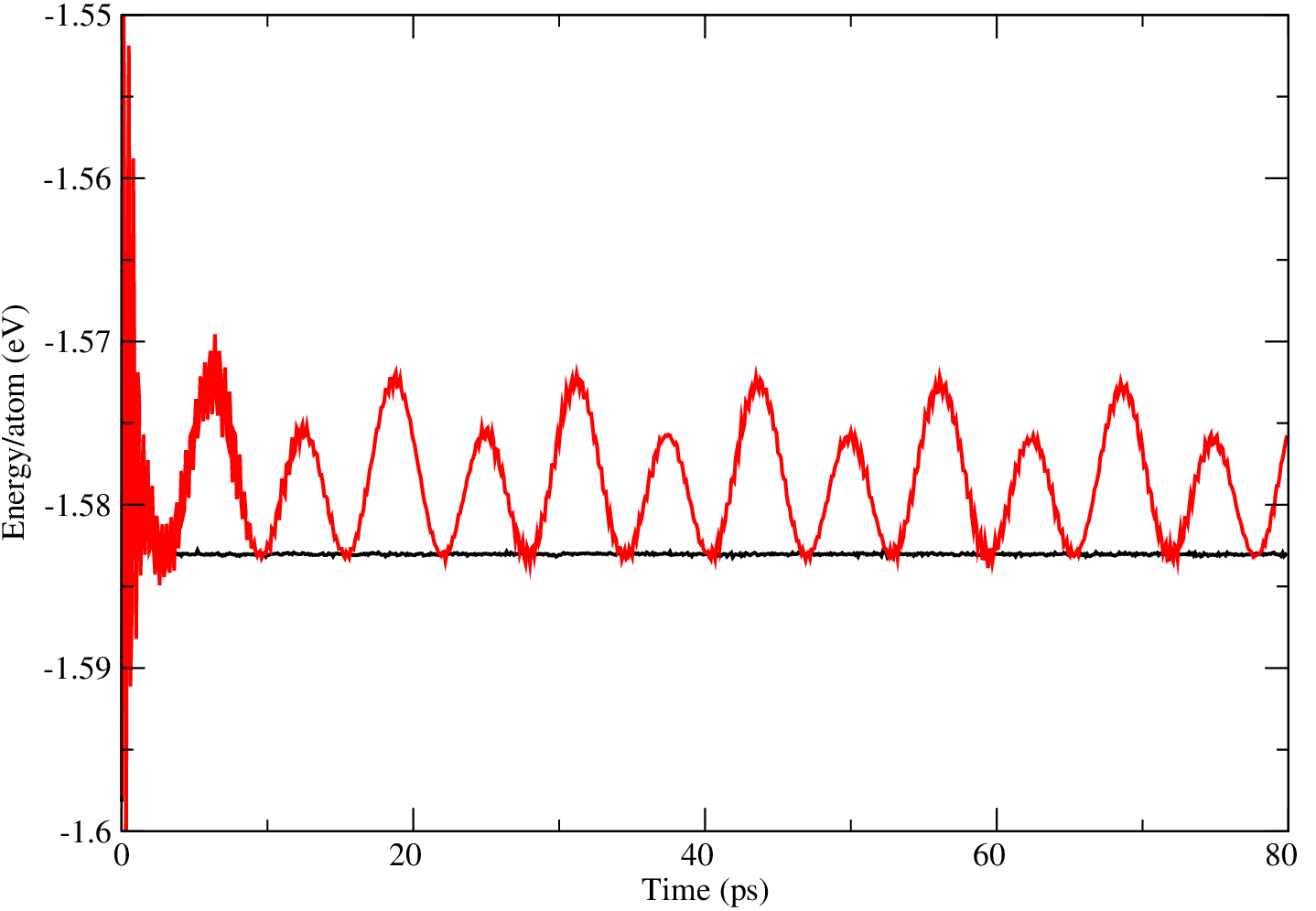}
\vspace{10mm} 

               \includegraphics[width=0.5\textwidth]{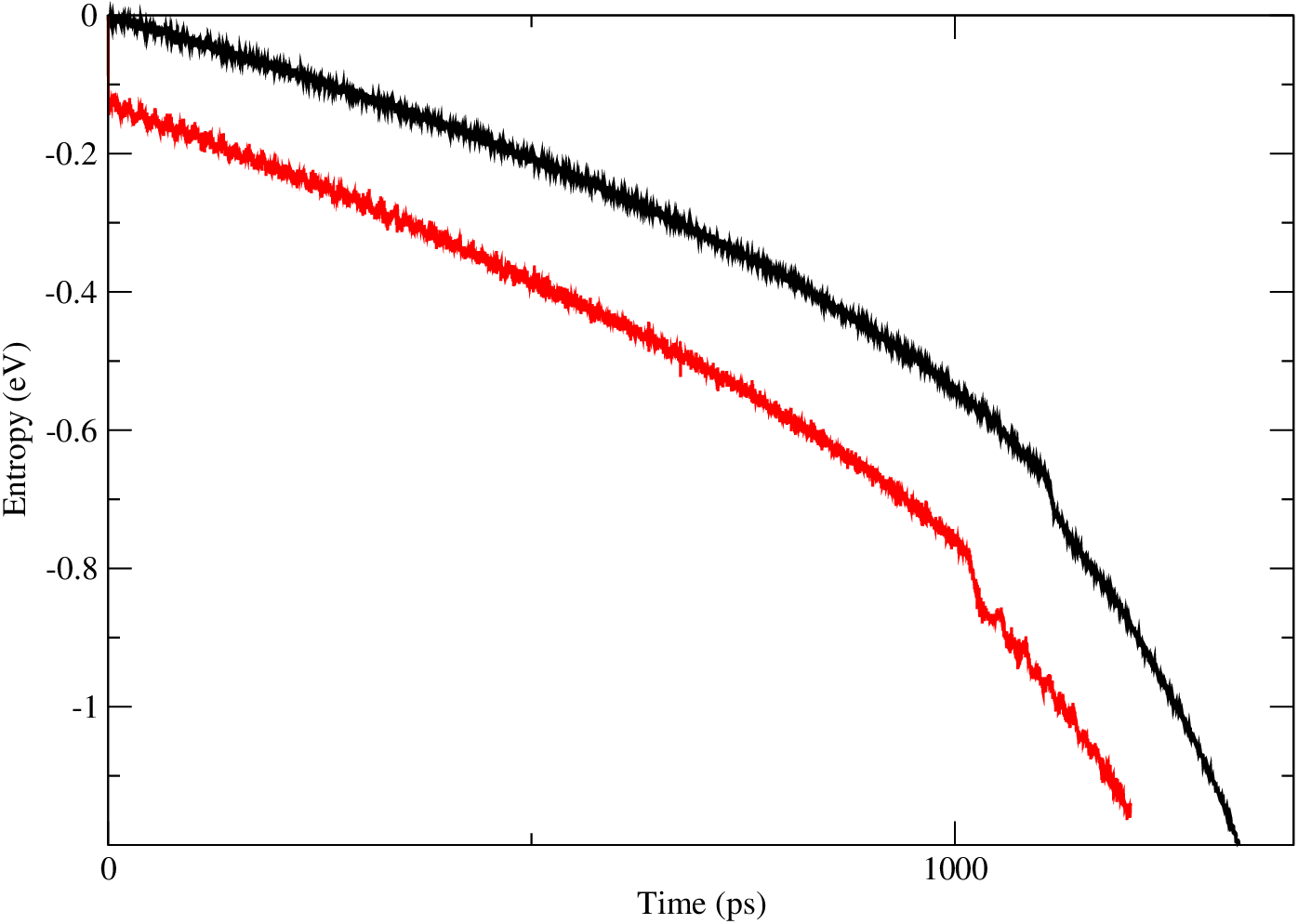}
         \caption{Volume, energy and entropy changes with time, red (grey)
           curves use the standard Parrinello Rahman method, black
           curves add algorithmic quench}
  \label{1}
\end{figure}

\begin{figure}[ht]
 \centering
                \includegraphics[width=0.8\textwidth]{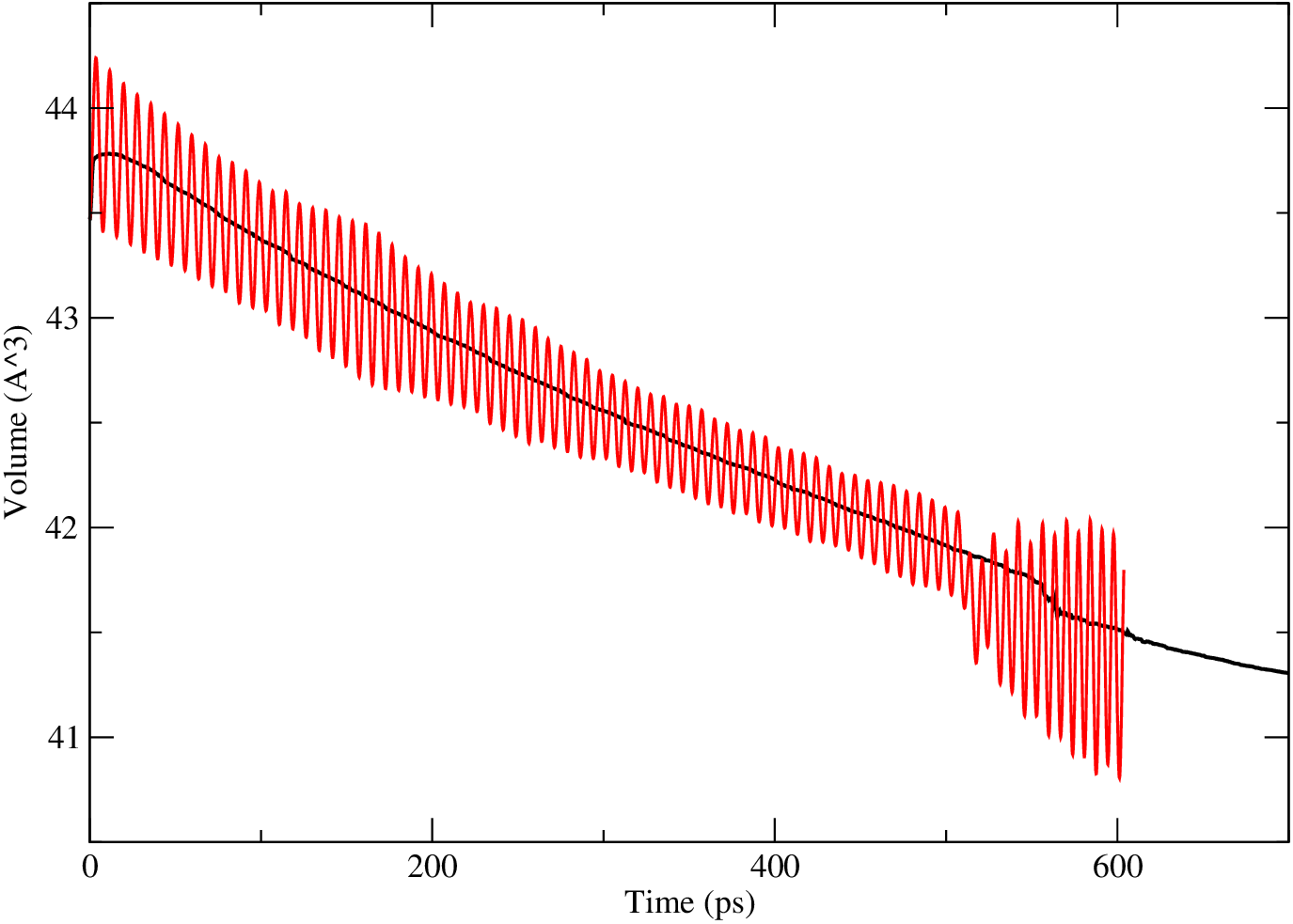}
\vspace{10mm} 

                \includegraphics[width=0.8\textwidth]{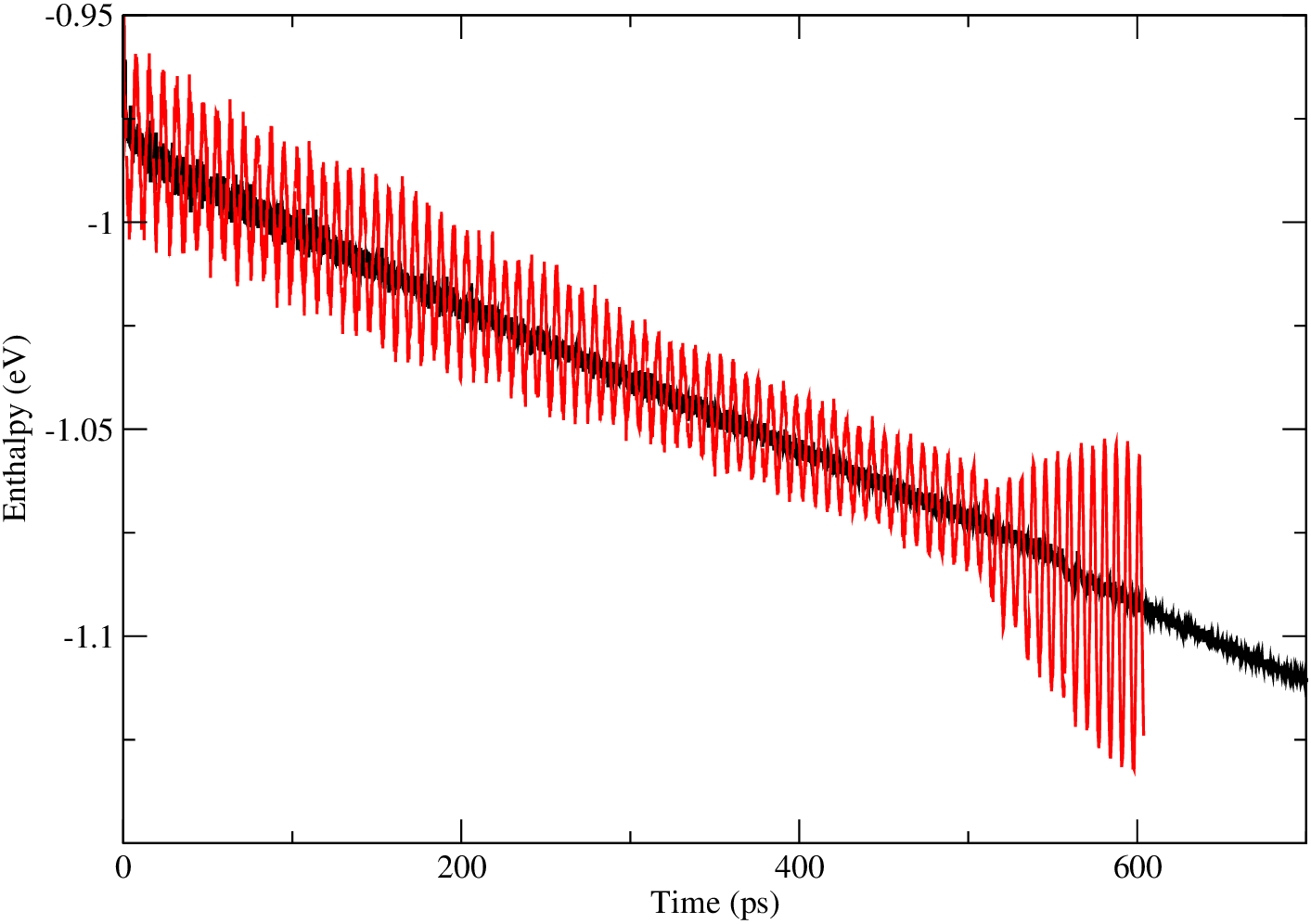}
         \caption{Volume and enthalpy changes with time, in a
           simulation cooling sodium from bcc to fcc simulations start
           at 500K and cool at 1.5ps/K, so that the transition after 500ps
occurs at about 150K. Red curves are the standard Parrinello Rahman
           method, black curves the same simulation adding algorithmic quench}
  \label{2}
\end{figure}

\end{document}